\documentclass[aps,prd,12pt,onecolumn,final,notitlepage,oneside,nobibnotes,
nofootinbib,superscriptaddress,showkeys,centertags]{revtex4-1}

\usepackage{amsmath,amsfonts,amssymb}
\usepackage{bm}
\usepackage{dcolumn}
\usepackage[dvips]{graphicx,graphics,color,epsfig}
\usepackage{wrapfig}
\usepackage{latexsym}
\usepackage{natbib}
\usepackage{verbatim}
\usepackage{bm}
\usepackage{tabulary}
\usepackage{natbib}
\usepackage{mathtext}
\usepackage[T2A]{fontenc}

\newcommand{\WID}[2]{#1_{\mathrm{#2}}}

\begin{document}
\title{Neutron Production by Cosmic-Ray Muons in Various Materials}

\author{\firstname{K.~V.}~\surname{Manukovsky}}
\affiliation{Institute for Nuclear Research, Russian Academy of Sciences,
pr. Shestidesyatiletiya Oktyabrya 7a, Moscow, 117312 Russia}
\affiliation{Institute for Theoretical and Experimental Physics, Bolshaya
Cheremushkinskaya 25, Moscow, 117218 Russia}
\author{\firstname{O.~G.}~\surname{Ryazhskaya}}
\affiliation{Institute for Nuclear Research, Russian Academy of Sciences,
pr. Shestidesyatiletiya Oktyabrya 7a, Moscow, 117312 Russia}
\author{\firstname{N.~M.}~\surname{Sobolevsky}}
\affiliation{Institute for Nuclear Research, Russian Academy of Sciences,
pr. Shestidesyatiletiya Oktyabrya 7a, Moscow, 117312 Russia}
\affiliation{Moscow Institute of Physics and Technology (State University),
Institutskii Lane 9, Dolgoprudnyi, Moscow Region, 141700 Russia}
\author{\firstname{A.~V.}~\surname{Yudin}}
\email{yudin@itep.ru}
\affiliation{Institute for Nuclear Research, Russian Academy of Sciences,
pr. Shestidesyatiletiya Oktyabrya 7a, Moscow, 117312 Russia}
\affiliation{Institute for Theoretical and Experimental Physics, Bolshaya
Cheremushkinskaya 25, Moscow, 117218 Russia}

\begin{abstract}
The results obtained by studying the background of neutrons produced by cosmic--ray muons in
underground experimental facilities intended for rare--event searches and in surrounding rock are presented.
The types of this rock may include granite, sedimentary rock, gypsum, and rock salt. Neutron production
and transfer were simulated using the Geant4 and SHIELD transport codes. These codes were tuned
via a comparison of the results of calculations with experimental data, in particular, with data of the
Artemovsk research station of the Institute for Nuclear Research (INR, Moscow, Russia) as well as via
an intercomparison of results of calculations with the Geant4 and SHIELD codes. It turns out that the
atomic--number dependence of the production and yield of neutrons has an irregular character and does not
allow a description in terms of a universal function of the atomic number. The parameters of this dependence
are different for two groups of nuclei: nuclei consisting of alpha particles and all of the remaining nuclei.
Moreover, there are manifest exceptions from a power--law dependence, for example, argon. This may
entail important consequences both for the existing underground experimental facilities and for those under
construction. Investigation of cosmic--ray--induced neutron production in various materials is of paramount
importance for the interpretation of experiments conducted at large depths under the Earth's surface.
\end{abstract}


\maketitle

\section*{INTRODUCTION}
Searches for rare events, including neutrino signals
from collapsing stars, neutrino oscillations, proton
decay, and traces of dark--matter particles, are
being performed at underground experimental facilities.
The problem of the background generated by
natural rock radioactivity and by cosmic--ray muons
is a key problem in such experiment. High--energy
muons penetrate easily through rock to large depths,
reaching underground facilities and producing neutrons
in direct interactions with nuclei and, starting
from a depth of several hundred meters, in hadron
and electromagnetic showers. It is noteworthy that
neutron production may proceed both in the facility
itself and in surrounding structures, shielding materials,
and rock. A background produced in rock is especially
hazardous since this may imitate rare
events: high--energy neutrons produced in such processes
and moderated in rock may go far away from
the track of the parent muon, becoming ``isolated'' \cite{01, 02}.
Coincidence schemes are then unable to remove
captures of such neutrons in the active zone of the
facility, and this may distort strongly experimental
results.


\begin{wrapfigure}{l}{0.55\textwidth}
\epsfxsize=0.55\textwidth
\vspace{-1.2cm}\center\epsffile{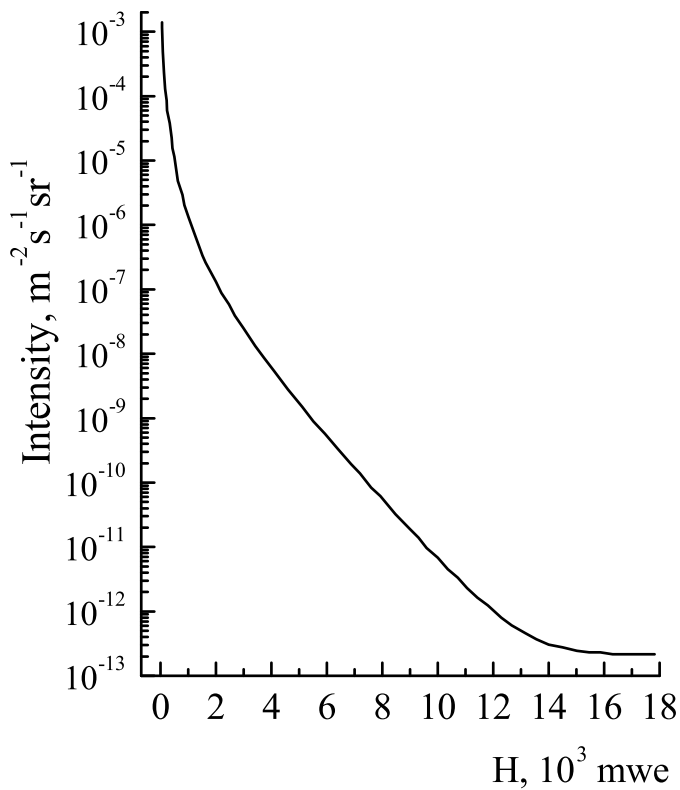}
\vspace{-1.0cm}\caption{\rm Intensity of muon flux as a function of depth.}
\label{Fig.muIntensity}
\end{wrapfigure}
The originally prevalent opinion was that, as
disposition of experiments becomes deeper, the contribution
of cosmic-ray muons to the formation of the
background should decrease substantially in proportion
to the reduction of the muon--flux intensity (see
Fig.~\ref{Fig.muIntensity}), but this proved to be untrue. The decrease
in the muon--flux intensity is accompanied by the
increase in the mean energy of muons involved. This
increase terminates only at a depth of about 10 kmwe
at the mean muon energy of about 430 GeV. The important
role of processes involving the production of
nuclear--active particles in hadron showers formed in
muon interactions with nuclei of surrounding matter
and the need for taking such processes into account
were first highlighted in \cite{03}. The $\mu+A \rightarrow \mu+n\pi+\chi$
reaction induced by deep--inelastic interactions of
muons $\mu$ with rock nuclei $A$ leads to the breakup
of these nuclei to fragments $\chi$ and to multiparticle
production of pions $\pi$, and this gives rise to a further
development of the nuclear cascade. At the depth
of 4000 mwe, for example, the disregard of this
process leads to a neutron-background intensity
underestimated by a factor of 2.5 \cite{04}. At large depths,
electromagnetic showers are yet another important
source of neutrons.

\begin{figure}[htb]
\epsfxsize=0.7\textwidth
\center\epsffile{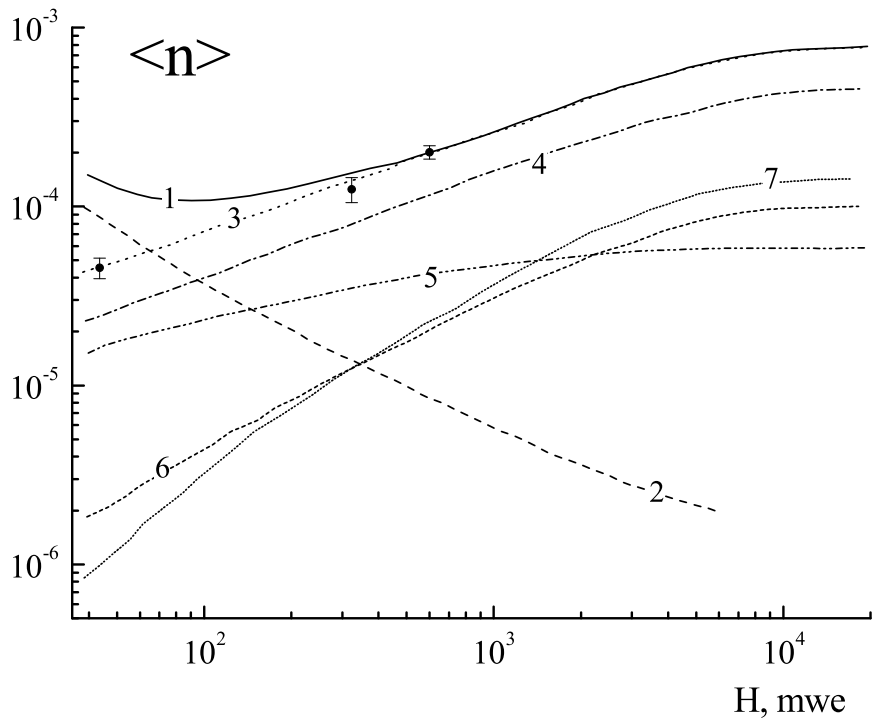}
\caption{\rm Number of product neutrons in 1 g $\mbox{cm}^{-2}$ of rock per muon as a function of depth: (curves) contributions of various
processes (see main of the text) and (points) results of the experiments under discussion.}
\label{Fig.neutr.gen.proc}
\end{figure}
The contributions of various
processes to neutron production per muon are shown
in Fig.~\ref{Fig.neutr.gen.proc} versus depth measured from the boundary of
the Earth's atmosphere. In this figure, curve 1 represents
the total number of neutrons produced in all
processes. At moderately small depths (up to about
80 mwe), $\mu^{-}$ capture (curve 2 in Fig.~\ref{Fig.neutr.gen.proc}) is a dominant
source of neutrons. At large depths, neutrons are
predominantly produced in nuclear showers (curve 4)
initiated by pions originating from the deep--inelastic
interaction of muons with surrounding--matter nuclei
and from electromagnetic showers generated by delta
electrons, bremsstrahlung photons, and electron--positron
pairs (curves 5, 6, and 7, respectively).
Curve 3 stands for the total number of neutrons produced
in all processes with the exception of the $\mu^{-}$ --
capture process. Figure~\ref{Fig.neutr.gen.proc} shows that, at depths larger
than 1000 mwe, the contribution of nuclear showers
to neutron production exceeds the contribution of
electromagnetic showers by a factor of 1.5.

The points with error bars in Fig.~\ref{Fig.neutr.gen.proc} represent data
from the experiments discussed below and performed
at the rock thicknesses of 25, 316 and 570 mwe and
the mean muon energies of 16.7, 86, and 125 GeV,
respectively.

\section*{DEBUGGING THE GEANT4 CODE PACKAGE}
In the calculations described below, the results
were obtained by using the Geant4 code package
in version 9.4 (patch 2) \cite{05}. This code package
permits detailed Monte Carlo calculations of particle
propagation through matter, possesses means
necessary for visualizing objects of complicated geometric
shapes, and incorporates a broad set of theoretical
models describing particle interactions with
matter. Particular attention was given to the choice
of physical models that are necessary for accurately
describing neutron production and propagation. The
set of neutrons originating from the interaction of
cosmic--ray muons with matter nuclei can be broken
down into four groups \cite{06}. Primary neutron originate
directly muon--nucleus interaction (upon the capture
of a negatively charged muon and muon spallation),
while secondary neutrons arise in nuclear and electromagnetic
showers. At small depths of up to 100 mwe,
neutrons produced upon muon capture are dominant,
while, at large depths (a few thousand mwe units)
hadron and electromagnetic showers are the main
source of neutrons.

Thus, many mechanisms are involved in the
neutron-production process. In view of this, it is
necessary to employ a broad set of physical models
that describe particle--interaction processes in various
energy ranges lying between a few eV units and several
hundred GeV units. The Geant4 code package
incorporates several standard versions of tuning of
physical models. In our calculations, we employ our
own set of models, relying on the analysis reported
in \cite{07}.

Let us mention briefly some important special features
of our approach. Final states of nuclei in photonuclear
interaction with a muon are described on
the basis of the chiral--phase--space (CHIPS) model
for photon energies below 3 GeV; at higher energies,
use is made of the quark--gluon string (QGS) model.
We apply the QGS model in describing nuclear interactions
at high energies (above 12 GeV) and the
binary--intranuclear--cascade (BiC) model at low energies
(below 6 GeV for neutrons and protons and
below 1.5 GeV for pions) and employ the low--energy
parametrized (LEP) model in the case of reactions in
the range of intermediate energies. One can observe
manifestations of matching of the different physical
models in Fig.~\ref{Fig.p.n.lead} below: the calculated value of the specific
neutron yield in a lead target is 15\%
lower than experimental
data at a proton energy of about 10 GeV (see
also the relevant discussion below). In order to remove
the excitation of the resulting nucleus, we used
models that take into account evaporation, fission,
Fermi breakup, and multifragmentation (in the case
of highly excited nuclei) processes. A model based on
the tables of experimentally measured cross sections
from the ENDF/B--VI database is used to describe
the transport of low--energy neutrons (whose energy
does not exceed 20 MeV) with allowance for elastic
and inelastic scattering, capture, and nuclear fission.

All of the results of our simulation that are presented
below were obtained on the basis of the same
set of physical models.

\section*{COMPARISON WITH EXPERIMENTAL DATA}
In order to control the accuracy of our calculations
and the debugging of the Geant4 code package, we
perform a series of test calculations and comparisons
of calculated data with known experimental results.

\begin{wrapfigure}{r}{0.4\textwidth}
\epsfxsize=0.4\textwidth
\vspace{-2.cm}\center\epsffile{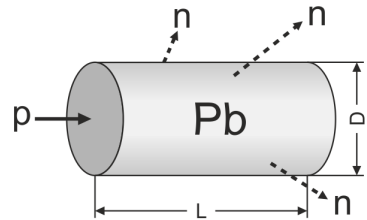}
\vspace{-1.6cm}\caption{\rm Cylindrical lead target of diameter $D$ and
length $L$}
\label{Fig.cylinder}
\end{wrapfigure}

As a starting point, we address the simple case where
one irradiates a cylindrical lead target of specific size
with protons of various energy (Fig.~\ref{Fig.cylinder}) and perform
a simulation of the neutron yield from it for this case
known from the literature. Experimental data and the
results of our calculations with the aid of the Geant4
code package and calculations of other authors from
the review article of Barashenkov~\cite{08} are given in Table~\ref{Tabl.1}, where good agreement between the calculated
and experimental data can be seen.

\begin{table}[htb]
\vspace{-0.2cm} \centering \caption{Mean number of neutrons escaping from a cylindrical
target of diameter $D$ per primary proton of energy $\WID{E}{p}$
(the target length is $L = 61$ cm)}\label{Tabl.1}
 \begin{tabular}{| c | c | c | c | c |}\hline\hline
 & & \multicolumn{3}{ c |}{Number of neutrons} \\ \cline{3-5}
  & & cal\-cu\-la\-ti\-ons & ex\-pe\-ri\-men\-tal  & Geant4--based  \\
 $D$, cm & $\WID{E}{p}$, GeV & \cite{08} & data~\cite{08} & cal\-cu\-la\-ti\-ons \\ \hline
 10.2 & 0.47 & $7.8 \pm 0.3$ & $8 \pm 0.4$  & 6.97 \\
      &    &   &     $6.4 \pm 0.3$ & \\\hline
  10.2 & 0.96 & $17.8 \pm 0.6$ & $16.6 \pm 0.8 $  & 16.65 \\
   &  &  &  $16.8 \pm 0.5 $  &  \\
   &  &  &   17.7 &  \\ \hline
  10.2 &  1.47 &  $25.1 \pm 0.1$ & $26.4 \pm 1.3$  & 25.1 \\
   &   &   &  $27.5 \pm 0.6$  &  \\
    &   &   &  29.4 &  \\ \hline
  20.4 & 0.47 &  $8.1 \pm 0.3$  & $8.7 \pm 0.4$ & 7.92  \\
   &  &   9.2 &  &   \\\hline
  20.4 & 0.96 &  $21.7 \pm 0.8$  & $20.3 \pm 1.1$ & 20.36 \\
   & &  22.2 &  &  \\ \hline
  20.4 & 1.47 & $31.5 \pm 1.2$ & $31.5 \pm 1.6$ & 31.84\\ \hline
 \end{tabular}
\end{table}

\begin{figure}[h!tb]
\begin{minipage}{0.46\textwidth}
\epsfxsize=1.\textwidth
\center\epsffile{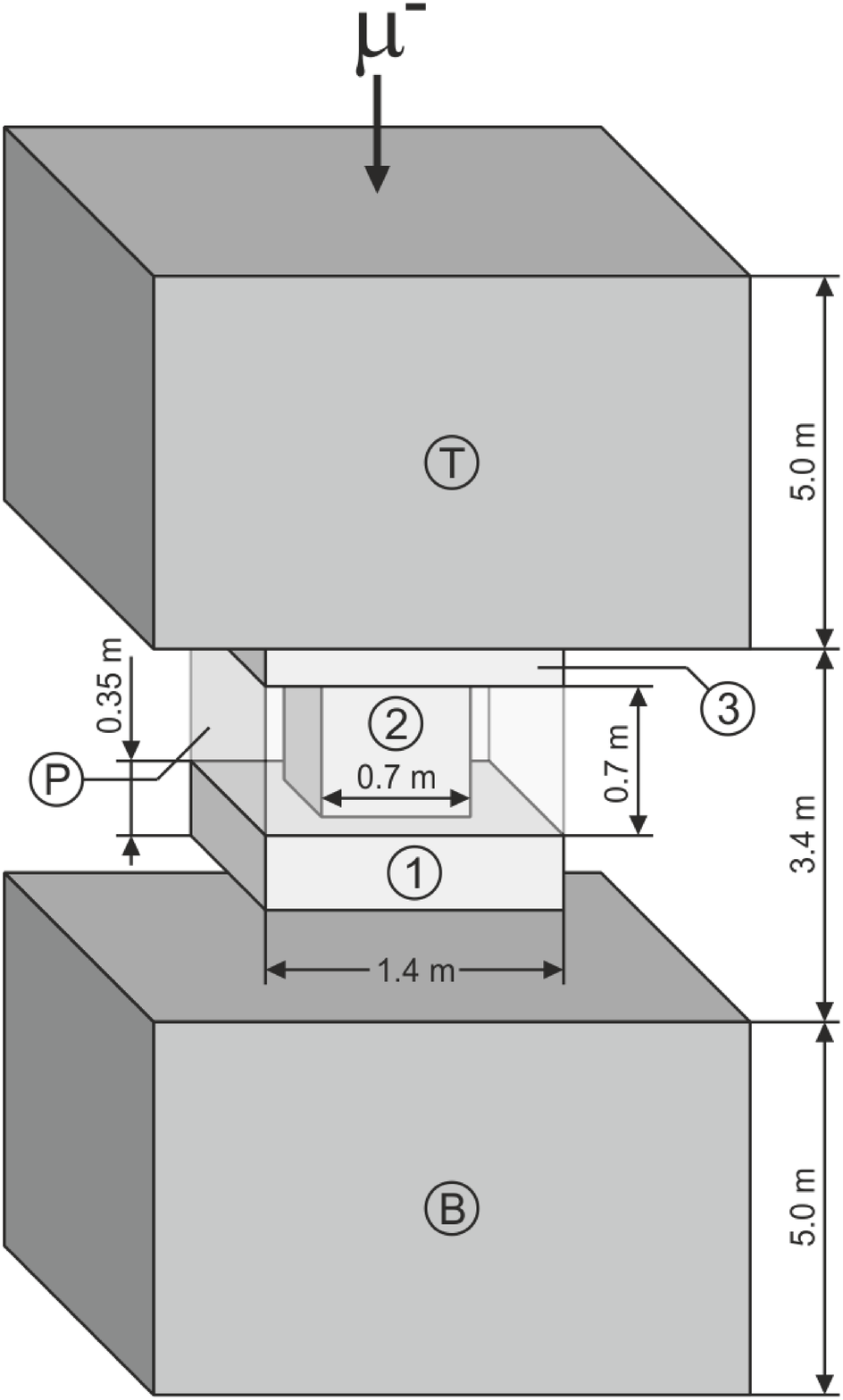}
\caption{\rm Layout of the setup for experiments at the depths
of 25 and 316 mwe. Volumes T and B simulate surrounding
rock, while volumes 1--3 represent the scintillator
used; P stands for surrounding paraffin.}
\label{Fig.ustanovka1}
\end{minipage}
\hspace{0.04\textwidth}
\begin{minipage}{0.46\textwidth}
\epsfxsize=1.\textwidth
\center\epsffile{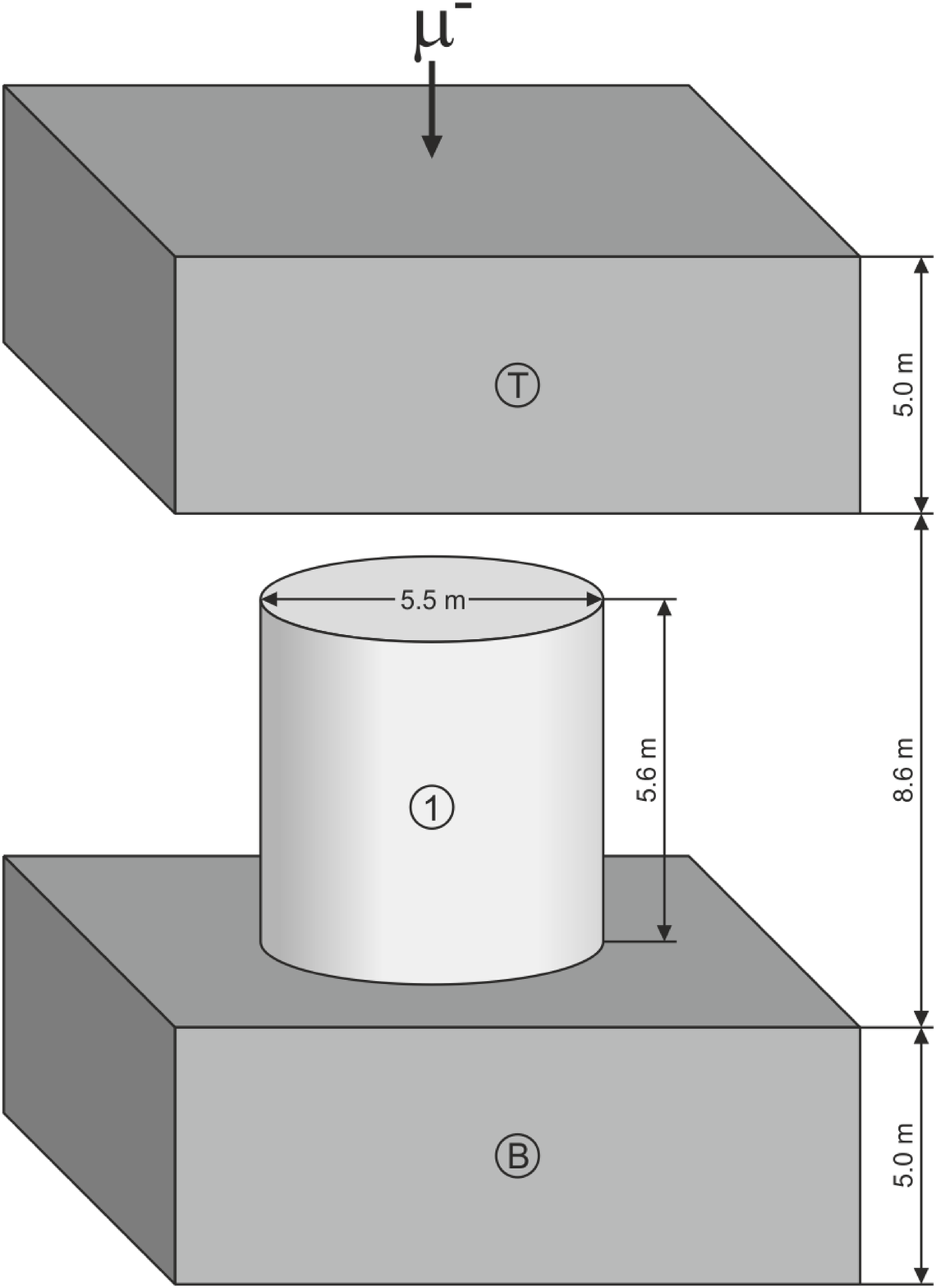}
\caption{\rm Layout of the setup for experiments at the depth of
570 mwe. Volumes T and B simulate surrounding rock,
while volume 1 stands for the scintillator used.}
\label{Fig.ustanovka2}
\end{minipage}
\end{figure}

We also perform a test simulation of the propagation
of cosmic--ray muons for various experimental
facilities of the Artemovsk research station of the
Institute for Nuclear Research (INR, Moscow, Russia)
\cite{01, 09, 10} that involve detectors based on a liquid
scintillator: specifically, in a gypsum mine at a
depth of 25 mwe \cite{10} and in rock--salt mines at the
depths of 316 and 570 mwe \cite{10, 11} (see also Fig.~\ref{Fig.neutr.gen.proc}).
The layout of the first facility for experiments at the
depths of 25 and 316 mwe is shown in Fig.~\ref{Fig.ustanovka1}. The
surrounding rock is represented by two plane layers
5 m thick above and below the facility, as can be seen
in Fig.~\ref{Fig.ustanovka1} (volumes T and B, respectively). Fixed energy
muons are assumed to travel vertically downward
at the facility center. Of the three rectangular
scintillation volumes (1---3), only volume 2 serves for
data acquisition, while the remaining two accomplish
event selection. Moreover, the central volume (2) is
surrounded by paraffin (P). For a muon effect to be
included in the accumulated data sample, the respective
muon should deposit at least 55 MeV in each of
volumes 1 and 3 and not less than 110 MeV in the
detecting volume (2).

The layout of the facility for experiments at the
depth of 570 mwe is shown in Fig.~\ref{Fig.ustanovka2}. Here, the
detecting scintillation volume is a cylindrical (1), and
the event-selection criterion requires that a muon
deposit not less than 1 GeV in this volume.

Table 2 gives experimental and calculated values
of the neutron-production yield per 1 g/cm$^2$,
$\WID{Y}{n}\equiv \WID{N}{n}/\left(\rho N_\mu l_\mu\right)$,
over the volume of the detecting scintillation
counter. Here, $\WID{N}{n}$ is the number of product
neutrons, $N_\mu$ is the total number of muons that traversed
the facility and which passed selection criteria,
and $l_\mu$ is muon range in matter of density $\rho$. From
Table 2, one can see that the calculated and experimental
data are very close to each other. In order to
assess the effect of the rock surrounding the experimental
facility on the number of product neutrons,
we perform a series of calculations, replacing the substance
surrounding the detector by various materials,
such as rock salt, scintillator, or gypsum. Table 2
shows that this effect is rather weak and becomes
noticeable at the minimum value of the muon energy.
The number of neutrons captured in the volume of the
same scintillation counter exhibits a similar behavior.
Thus, the presence of rock has but a slight effect on
the total of product neutrons detected by the counter.

\begin{table}[htb]
\vspace{-0.2cm} \centering \caption{Comparison of the experimental and calculated values
of the neutron--production yield $\WID{Y}{n}$ at various depths for
various materials surrounding scintillation counters
(in $n/\mu/(\mbox{g}/\mbox{cm}^2)$ units). LS denote a liquid scintillator.}\label{Tabl.2}
 \begin{tabular}{| c | c | c | c | c | c | c |}\hline\hline
 & \multicolumn{2}{ c |}{Experiment} & \multicolumn{4}{ c |}{Calculation}\\ \cline{2-7}
 Depth, &  &  &  & \multicolumn{3}{ c |}{$\WID{Y}{n}$} \\ \cline{5-7}
  mwe & $E_\mu$, GeV  & $\WID{Y}{n}$ & $E_\mu$, GeV & rock salt & LS & gypsum \\ \hline
 25 & 16.5 $\pm$ 8.1 & 0.36 $\pm$ 0.03 \cite{12} & 16.8 & 0.388 & 0.55 & 0.392 \\
    & 16.7 $\pm$ 8.2 &  0.47 $\pm$ 0.05 \cite{10} &  &  &  &  \\ \hline
 316 &  86 $\pm$ 18 & 1.2 $\pm$ 0.12 \cite{10} & 86 & 1.2 & 1.26 & 1.28 \\ \hline
 570 & 125 $\pm$ 22 & 2.04 $\pm$ 0.24 \cite{11} & 125 & 1.71 & 1.64 & --- \\ \hline
 \end{tabular}
\end{table}

\section*{COMPARISON OF THE RESULTS BASED ON THE APPLICATION OF THE SHIELD AND GEANT4 CODES}
In order to calibrate Geant4 debugging, we compare
the results of our calculations for a test problem
with the results reported in \cite{13} and obtained by using
the SHIELD transport code \cite{14}, which proved to be
successful in various applications. As a test problem,
we once again consider the problem of the neutron
yield from a lead target irradiated with a proton beam
whose energy varies over a broad range. The target
dimensions are $L = 60$ cm in length and $D = 20$ cm
in diameter.
\begin{figure}[htb]
\epsfxsize=0.8\textwidth
\center\epsffile{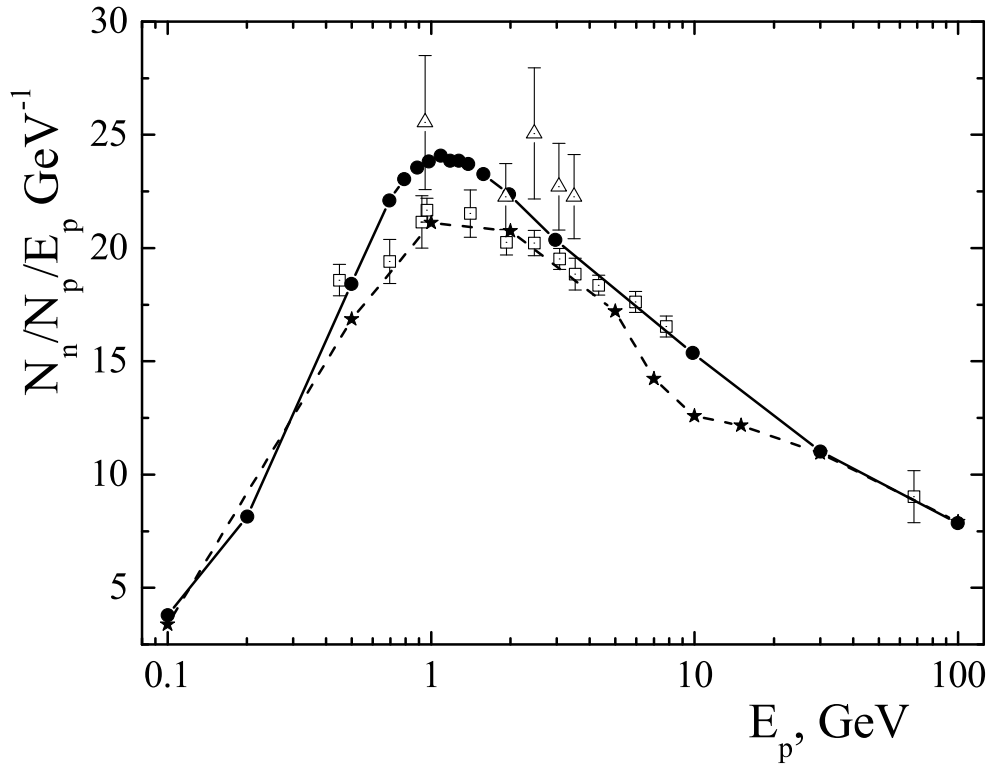}
\caption{\rm Specific yield of neutrons from a cylindrical lead
target as a function of proton energy. The displayed
points stand for experimental data. The solid and dashed
curves represent the results of the calculations performed
by using the SHIELD and Geant4 codes, respectively.}
\label{Fig.p.n.lead}
\end{figure}
The results are shown in Fig.~\ref{Fig.p.n.lead}, where
the quantity $\WID{N}{n}/(\WID{N}{p}\WID{E}{p})$ is plotted versus energy.
Here, $\WID{N}{n}$ is the number of neutrons that escaped from
the target volume and $\WID{N}{p}$ and $\WID{E}{p}$ are, respectively, the
number of protons and their energy in GeV units. The
open symbols with error bars stand for experimental
data from various sources. The solid and dashed
curves represent the results based the application of,
respectively, the SHIELD and Geant4 codes. The
uppermost group of five experimental points represented
by open triangles is worthy of special note.
These are data from \cite{15}, and a point from the lowest
group (boxes) corresponds to each point in the uppermost
group. They were obtained within the same
experiment but by different methods. The reason
behind this difference is not clear; therefore, there
are doubts about the reliability of these five points.
With allowance for the experimental uncertainties,
the two transport codes yield anyway mutually consistent
results, which agree with experimental data.
As was mentioned above, the discrepancy with experimental
data in Geant4--based calculations around
$\WID{E}{p} \sim 10$ GeV is due to the absence of an adequate
model for describing interactions in the energy range
between the ranges of applicability of the BiC and
QGS theoretical models. Among all possibilities that
are available within Geant4 at the present time and
which we considered here, the use of the LEP model
provides results that are the closest to experimental
data in the energy range between 1.5 and 12 GeV, but
which are by no means precise.

\section*{NEUTRON PRODUCTION BY MUONS}
\subsection*{Dependence on the Atomic Weight}
\begin{wrapfigure}{r}{0.4\textwidth}
\epsfxsize=0.4\textwidth
\vspace{-2.cm}\center\epsffile{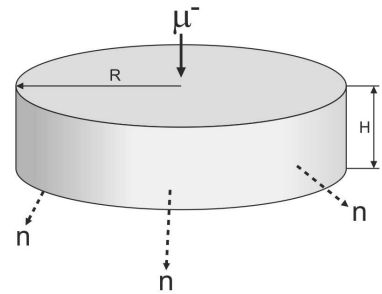}
\vspace{-1.6cm}\caption{\rm Scheme of a numerical experiment for determining
the dependence of the number of product neutrons on the
atomic weight of materials.}
\label{Fig.mu.cylinder}
\end{wrapfigure}

In order to determine the dependence of the number
of product neutrons on the atomic weight of the
material being considered, we perform a numerical
simulation of muon propagation for muons of various
energy through a target having a simple geometric
shape and featuring a uniform density distribution,
the size of the target being infinite in the direction
orthogonal to the muon velocity (see Fig.~\ref{Fig.mu.cylinder}). 

\begin{figure}[h!tb]
\begin{minipage}{0.48\textwidth}
\epsfxsize=1.\textwidth
\center\epsffile{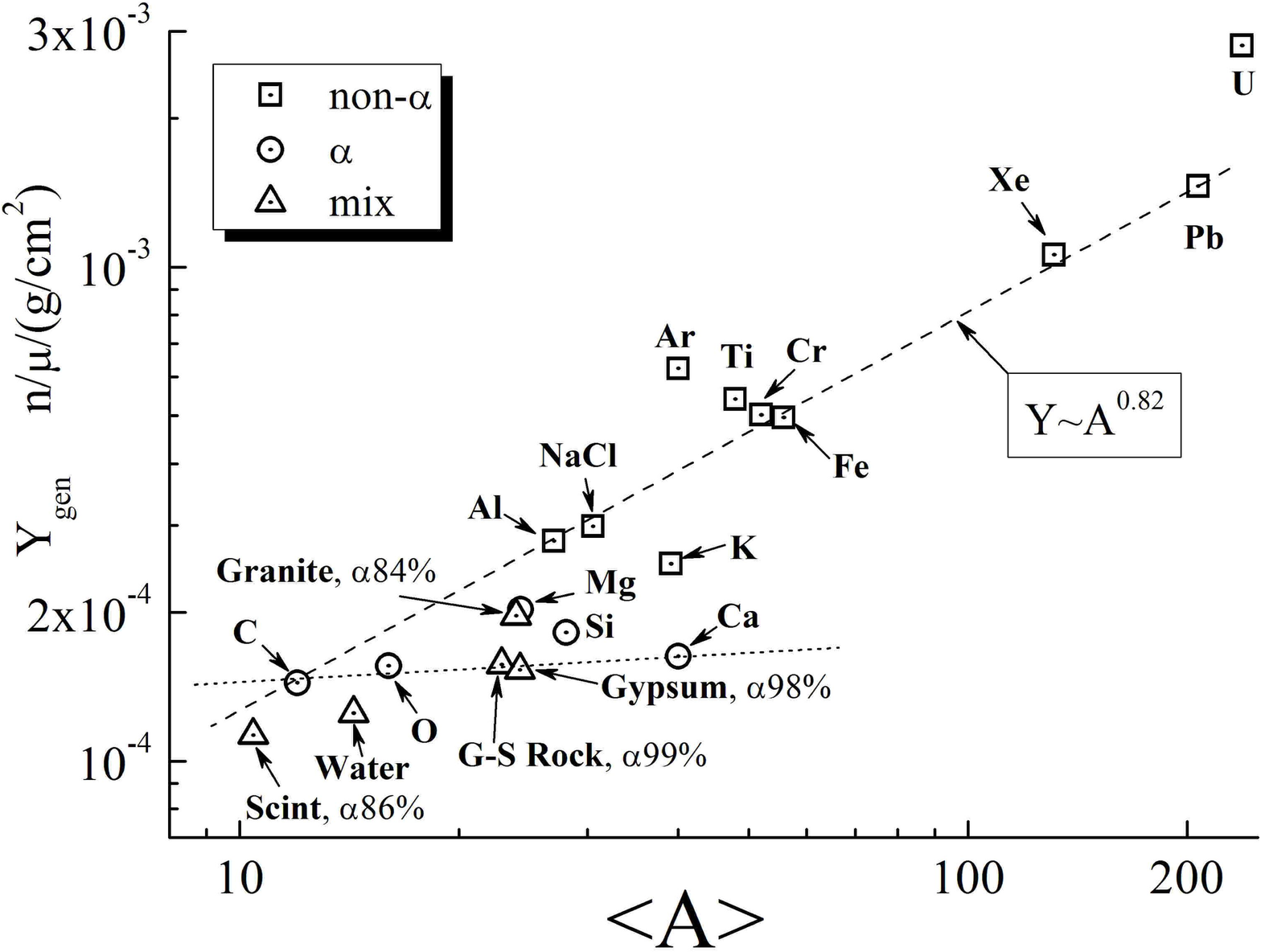}
\caption{\rm Neutron-production yield as a function of the atomic weight 
of matter for muons of energy 100 GeV.}
\label{Fig.Gen100}
\end{minipage}
\hspace{0.01\textwidth}
\begin{minipage}{0.48\textwidth}
\epsfxsize=1.\textwidth
\center\epsffile{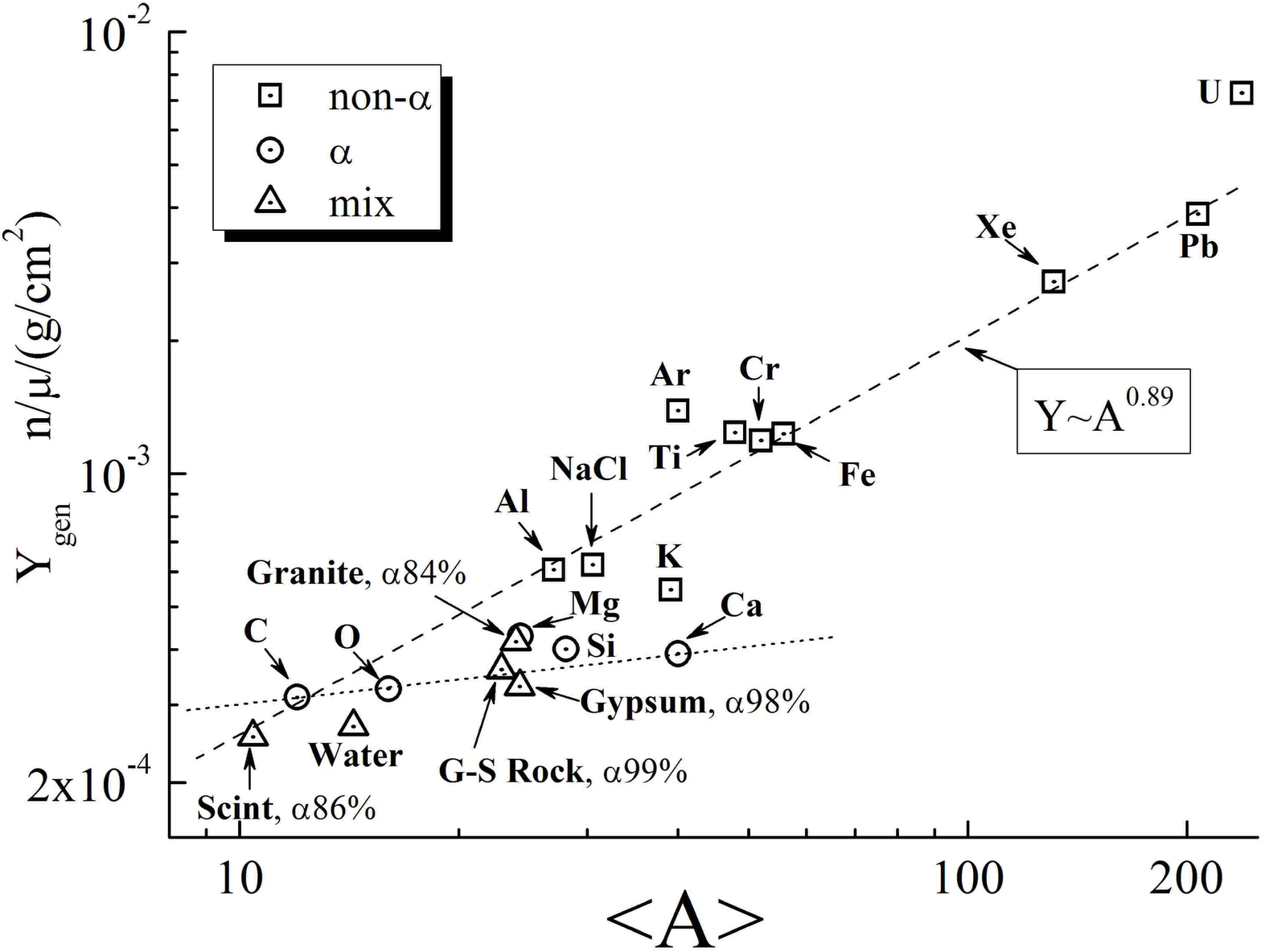}
\caption{\rm As in Fig.~\ref{Fig.Gen100}, but for the muon energy of 280 GeV. G-S denotes
the rock of the Gran Sasso underground laboratory.}
\label{Fig.Gen280}
\end{minipage}
\end{figure}

For each material, the total target thickness was chosen
to be equivalent to 2000 g cm$^{-2}$, which corresponds
to themean muon energy loss of about 5\% (for muons
of energy 100 GeV). The results of the calculation
for muons of energy 100 and 280 GeV are given in
Figs.~\ref{Fig.Gen100} and \ref{Fig.Gen280}, respectively. They show the neutron production
yield $\WID{Y}{n}$ [neutrons/muon/(g cm$^{-2}$)] as a
function of the mean atomic weight of the target material.
It should be noted that the data presented here
were obtained upon the event selection on the basis
of the spectrum of muon energy depositions: events
in which the muon energy deposition in the target
exceeded half of the most probable energy deposition
were selected for a comparison with experimental
data. A detailed analysis reveals that the selection
in question has but a slight effect on the production
yield and does not lead to qualitative changes in the
character of the resulting dependencies.

The calculated values of the number of product
neutrons can be broken down into three groups
markedly different in the target chemical composition:
those for targets free from alpha-particle nuclei (for
example, Fe or NaCl); those for targets from purely
alpha-particle nuclei (C, Ca, and so on); and those for
targets from materials of mixed composition (such as
the rock of the Gran Sasso underground laboratory),
in which case the fraction of alpha-particle nuclei is
indicated in the figure. In the first group, the neutron production
yield $\WID{Y}{n}(A)$ is well approximated by a
dependence of the form $\WID{Y}{n}(A)\propto A^k$, where $k \approx 0.82$
for $E_\mu=100$~GeV and $k \approx 0.89$ for $E_\mu=280$~GeV
(upper curves in Figs.~\ref{Fig.Gen100} and \ref{Fig.Gen280}). For alpha--particle
nuclei, the atomic--weight dependence of the number
of product neutrons is weak (lower curves). Finally,
the dependence for materials of mixed composition
exhibits an intermediate behavior between the
preceding two cases, the relative positions of the
curves correlating substantially with the fraction of
alpha-particle nuclei in the chemical composition of
the target. Nevertheless, some chemical elements
(for example, potassium and uranium) do not fit
satisfactorily in this pattern.

\begin{figure}[h!tb]
\begin{minipage}{0.48\textwidth}
\epsfxsize=1.\textwidth
\center\epsffile{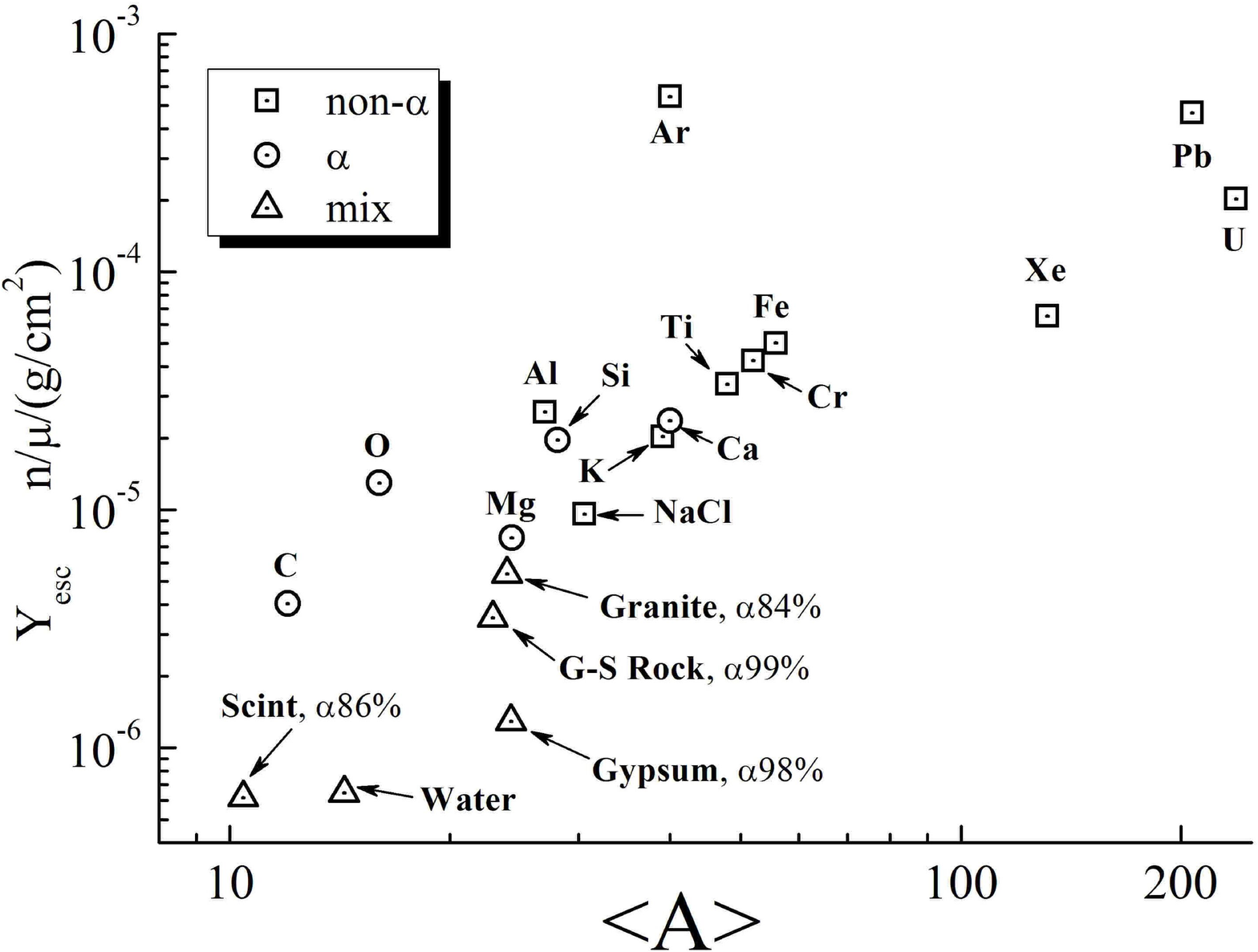}
\caption{\rm Neutron yield as a function of the atomic weight of materials for muons of energy 100 GeV.}
\label{Fig.Esc100}
\end{minipage}
\hspace{0.01\textwidth}
\begin{minipage}{0.48\textwidth}
\epsfxsize=1.\textwidth
\center\epsffile{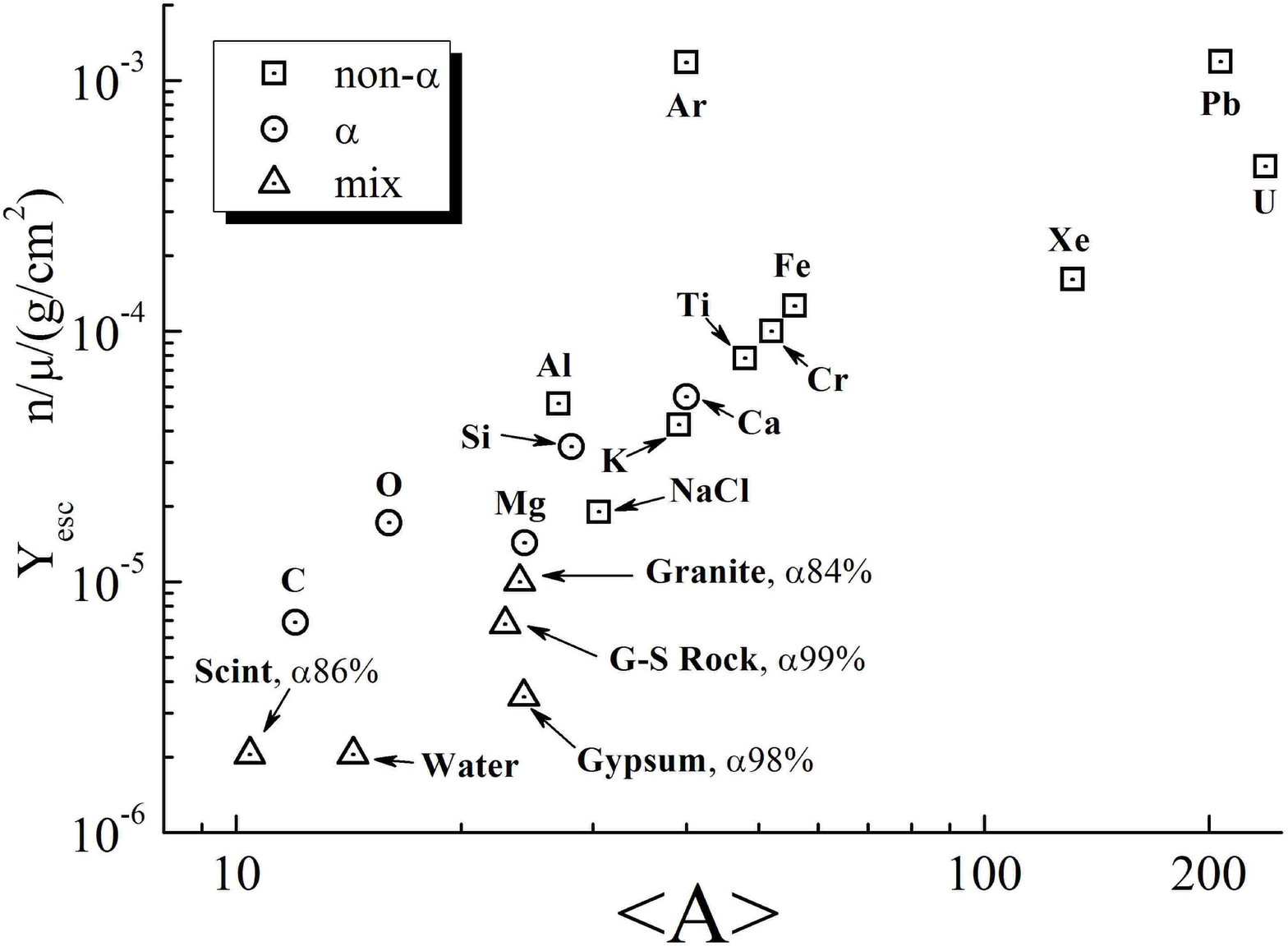}
\caption{\rm As in Fig.~\ref{Fig.Esc100}, but for the muon energy of 280 GeV.}
\label{Fig.Esc280}
\end{minipage}
\end{figure}

In addition to the number of product neutrons, the
calculations made it possible to determine the number
of neutrons that escaped from the target. The results
for muons of energy 100 and 280 GeV are shown
in Figs.~\ref{Fig.Esc100} and \ref{Fig.Esc280}. The neutron yield depends not
only on the atomic weight but also on many other
extra factors (such as the target size and shape and
the parameters of the neutrons-transfer process) and
therefore does not follow any simple dependence \cite{16}.
It follows that only via a numerical simulation can one
precisely estimate the neutron background generated
by cosmic--ray muons in modern underground detectors
of complicated design.

\subsection*{Dependence on the Muon Energy}
\begin{figure}[htb]
\epsfxsize=0.7\textwidth
\center\epsffile{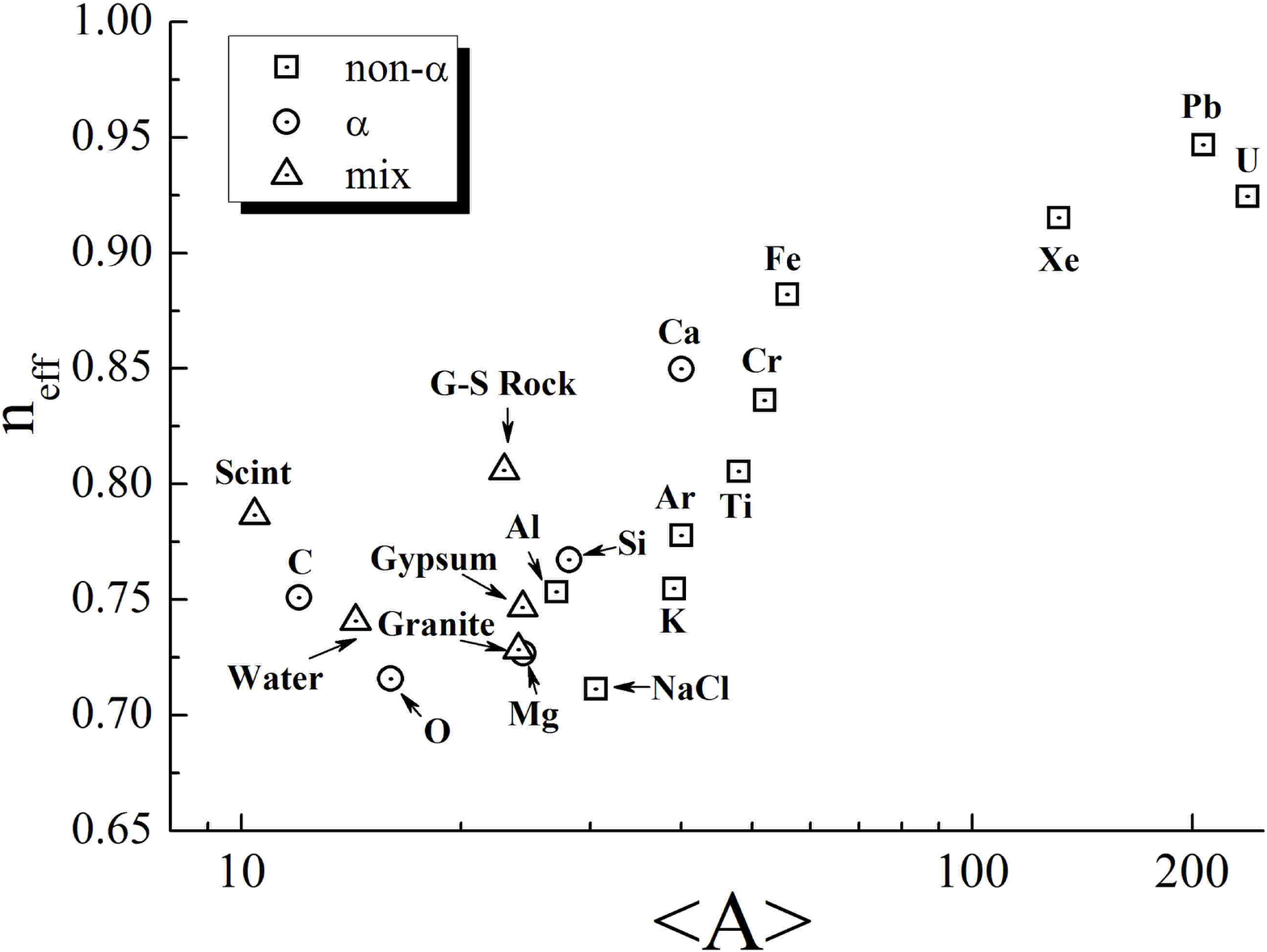}
\caption{\rm Effective exponent $\WID{n}{eff}$ in the muon--energy 
dependence of the neutron--production yield, $\WID{Y}{n}\sim E_{\mu}^{\WID{n}{eff}}$.}
\label{Fig.n.eff}
\end{figure}

The energy dependence of the neutron--production
yield is frequently represented in the power--law form
$\WID{Y}{n}\sim E_\mu^n$, where the exponent n is smaller than or on
the order of unity. Figure~\ref{Fig.n.eff} shows the values of the
effective exponent $\WID{n}{eff}$ that were obtained from a comparison
of the production yields at the muon--energy
values of 100 and 280 GeV for the same materials as
in the preceding figures. The majority of these values
fall within a rather narrow range, the average value
being 0.75. Alpha--particle nuclei show a moderate
muon--energy dependence; for other materials, the
exponent is somewhat higher, and there is a trend
toward an increase in $\WID{n}{eff}$ with increasing A. Uranium
and lead have a maximum value of $\WID{n}{eff}\lesssim 1$.

\subsection{Case of Argon}
\begin{wrapfigure}{r}{0.35\textwidth}
\epsfxsize=0.35\textwidth
\vspace{-2.cm}\center\epsffile{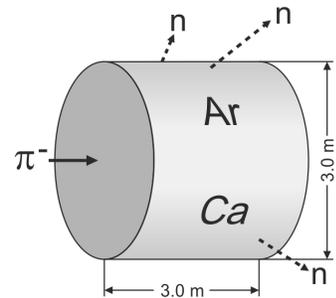}
\vspace{-1.6cm}\caption{\rm Scheme of a numerical experiment that employs
the SHIELD transport code.}
\label{Fig.pi.cylinder}
\end{wrapfigure}

Argon and calcium attract particular attention,
since these species, which have very close atomic
weights, differ strongly (several fold) in the number of
product neutrons. In order to study this effect, we
perform a calculation on the basis of the SHIELD
transport code for the problem of a cylindrical target
(its radius and length are $R = 1.5$ m and $L =
3$ m, respectively) irradiated along the axis with a
beam of negatively charged pions with energies of 1,
10, and 100 GeV (see Fig.~\ref{Fig.pi.cylinder}) 

Liquid argon ($\rho = 1.65$ g cm$^{-3}$) or calcium 
($\rho = 1.55$ g cm$^{-3}$) serve
as a target material. The neutron production and
neutron yield from the target are calculated. In the
case being considered, the former is the number of
neutrons characterized by energies below 14.5 MeV
and produced within the target in hadron-induced
nuclear reactions (they include cascade neutrons of
energy above 14.5 MeV). The latter is the number
of neutrons whose energy is below 14.5 MeV and
which go beyond the target boundary. The results are
presented in Table~\ref{Tabl.3}.
\begin{table}[htb]
\vspace{-0.2cm} \centering \caption{Comparison of the production and yield of neutrons
for a cylindrical target ($R = 1.5$ m and $L = 3$ m)
exposed to pions of various energy (liquid argon or calcium
serves as a target materia)}\label{Tabl.3}
 \begin{tabular}{| c | c | c | c | c |}\hline\hline
 $E_\pi$, & \multicolumn{2}{ c |}{Ar} & \multicolumn{2}{ c |}{Ca}\\ \cline{2-5}
 GeV & production & yield & production & yield\\ \hline 
 1 & 16.7 & 17.1 & 7.47 & 4.69\\ \hline
 10 & 77.0 & 78.8 & 33.6 & 20.7\\ \hline
 100 & 381 & 390 & 161 & 99.1\\ \hline
 \end{tabular}
\end{table}

The reasons behind the large distinction between
the results for $^{40}$Ar and $^{40}$Ca are the following. First,
the neutron multiplicity is higher in hadron--nucleus
reactions on argon. Indeed, the energies of separation
of a single neutron and a neutron pair from the $^{40}$Ar
nucleus are 9.87 and 16.5 MeV, respectively, but, for
$^{40}$Ca, the respective energies are 15.6 and 29.0 MeV
\cite{17}.

\begin{figure}[htb]
\epsfxsize=0.47\textwidth
\epsffile{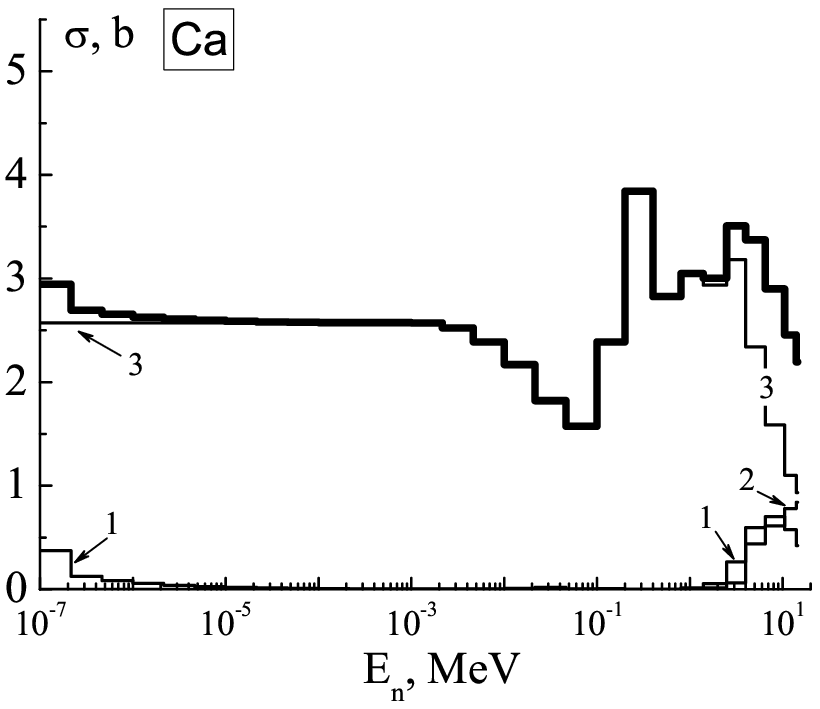}
\hspace{0.02\textwidth}
\epsfxsize=0.47\textwidth
\epsffile{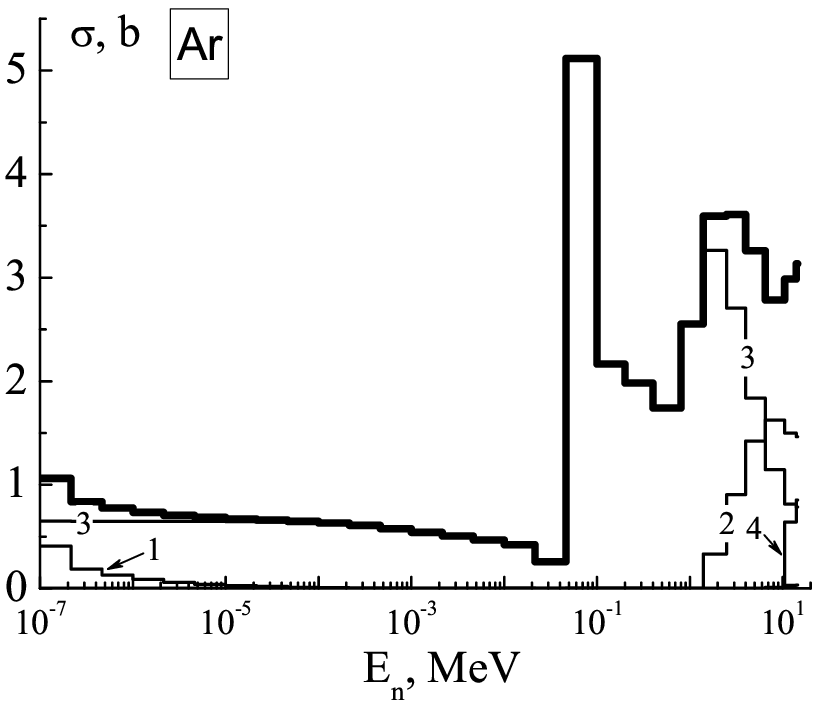}
\caption{\rm Comparison of the energy dependences of the group cross sections from [18] for neutron reactions in calcium and argon. The notation used is explained in the main body of the text.}
\label{Fig.Ar.Ca}
\end{figure}

The second reason is that the neutron cross sections
at energies below 14.5 MeV are markedly different
for these two nuclear species. Figure~\ref{Fig.Ar.Ca} shows
the cross sections for reactions in calcium and argon
versus the neutron energy $E_n$. The figures on the lines
mark the cross sections for (1) capture, (2) inelastic
scattering, (3) elastic scattering, and (4) (n, 2n)
reactions. The thick line represents the total cross.
In argon, the (n, 2n) reaction proceeds, but there is
no such reaction in calcium. Moreover, this figure
shows that the total cross section for neutrons of
energy below about 0.1 MeV in argon is three to
four times smaller than its counterpart for calcium.
Accordingly, the neutron ranges in argon are three
to four times longer than those in calcium, and this
facilitates neutron escape from the target volume.

\section*{CONCLUSIONS}
The importance of correctly taking into account
the background in experiments devoted to searches
for rare events is worth emphasizing in conclusion.
The effect of the irremovable (in principle) background
of neutrons produced by cosmic--ray muons
under the Earth's surface is treatable properly only if
we are able to calculate it correctly with allowance
for all special features of the experimental setup being
considered, the surrounding shielding elements, and
rock. The above example of the comparison of calcium
and argon shows that the use of oversimplified
model concepts of neutron--production processes in
various materials may lead to a significant (several fold)
deviations from true result. The foregoing is all
the more true since liquid argon, which is a strong
source of neutrons, is an indispensable ingredient
of many experiments conducted at the present time
(see, for example, \cite{19}). Thus, only a thorough
understanding of processes involving neutron production
by cosmic rays in various materials would
make it possible to interpret correctly the results
of experiments performed at large depths under the
Earth's surface.

\section*{ACKNOWLEDGMENTS}
This work was supported in part by the Russian
Foundation for Basic Research (project no. $15{-}02{-}01056\_a$) 
and was funded by a grant (no. 3110.2014.2)
for support of leading scientific schools. The financial
support within the program of basic investigations of
Presidium of Russian Academy of Sciences Fundamental
Properties of Matter and Astrophysics is also
gratefully acknowledged.

\end{document}